%% file: simd_mlssc.tex
\documentclass{article}
\usepackage{spconf}

\usepackage{cite}

\usepackage{ifpdf}
\ifpdf
\usepackage[final,pdfborder={0 0 0}]{hyperref}
\fi
\ifpdf
\hypersetup{pdfauthor={Pascal Giard, Gabi Sarkis, Claude Thibeault, and Warren J. Gross},%
  pdftitle={Fast Software Polar Decoders},
  colorlinks=true,
  linkcolor=blue,
  citecolor=blue
}%
\fi

\usepackage[cmex10]{amsmath}
\usepackage{amssymb}
\usepackage{bm}
\usepackage[varg]{txfonts}
\let\mathbb=\varmathbb
\DeclareSymbolFont{letters}{OML}{ztmcm}{m}{it}

\usepackage{fixltx2e}

\usepackage{graphicx}
\usepackage{placeins}
\usepackage{float}
\usepackage{tabularx,colortbl}
\usepackage{multirow}
\usepackage{dcolumn}
\usepackage{booktabs}
\usepackage{tikz}
\usepackage{pgfplots}
\usetikzlibrary{plotmarks}
\usetikzlibrary{shapes,positioning,arrows,decorations.markings,fit,calc,patterns}
\usepackage{subfig}

\usepackage{color}

\usepackage{siunitx} %
\usepackage{calc} %

\newcommand{\mvec}[1]{\bm{#1}}
\newcommand{\est}[1]{\hat{u}_{#1}}

\newcommand{\llr}[1]{\lambda_{#1}}
\newcommand{\sgn}[1]{\text{sign}(#1)}

\newcolumntype{C}[1]{>{\centering}m{#1}}

\begin{document}
\title{Fast Software Polar Decoders}

\name{Pascal Giard$^{\star \dagger}$, Gabi Sarkis$^{\star}$, Claude Thibeault$^{\dagger}$, and Warren J. Gross$^{\star}$}
\address{$^{\star}$McGill University, Montr\'eal, Qu\'ebec, Canada\\
  $^{\dagger}$\'Ecole de technologie sup\'erieure, Montr\'eal, Qu\'ebec, Canada}

\maketitle

\begin{abstract}
Among error-correcting codes, polar codes are the first to provably achieve channel capacity with an explicit construction. In this work, we present software implementations of a polar decoder that leverage the capabilities of modern general-purpose processors to achieve an information throughput in excess of 200 Mbps, a throughput well suited for software-defined-radio applications. We also show that, for a similar error-correction performance, the throughput of polar decoders both surpasses that of LDPC decoders targeting general-purpose processors and is competitive with that of state-of-the-art software LDPC decoders running on graphic processing units.

\end{abstract}


\section{Introduction}
\label{sec:intro}
Being the first error-correcting codes with an explicit construction to provably achieve the symmetric channel-capacity, polar codes have drawn significant attention since their introduction in \cite{Arikan2008,Arikan2009}. Many hardware implementations of the successive-cancellation (SC) algorithm were able to exploit the regular structure of polar codes to reduce implementation complexity \cite{Leroux2013,Mishra2012,Raymond2013}.

Recently, new decoding algorithms derived from SC were proposed with the explicit aim of increasing throughput without degrading error-correction performance: simplified successive-cancellation (SSC)\cite{Alamdar-Yazdi2011}, maximum-likelihood simplified successive-cancellation (ML-SSC)\cite{Sarkis2013}, two phase SC (TPSC)\cite{TWOPhasePamuk} and Fast-SSC\cite{Sarkis2013b}.

\textit{Contribution and Outline:} This paper presents fast software polar decoders with an information throughput that can exceed 200~Mbps using only one core of an Intel i7-2600 x86 CPU running at 3.4~GHz. To that end, Section~\ref{sect:polar} reviews polar codes, and Section~\ref{sect:RSM-SCC} briefly reviews the Fast-SSC decoding algorithm. Then, Section~\ref{sect:map} details different implementations of the algorithm on an x86 CPU featuring single-instruction-multiple-data (SIMD) capability. Section~\ref{sect:results} provides throughput results. Finally, Section~\ref{sect:conclusion} concludes this paper.

\section{Polar Codes}
\label{sect:polar}

Polar codes are constructed recursively by applying linear transformations that create $N$ channels where, as $N \rightarrow \infty$, the probability of an error in transmission of a subset of the $N$ channels tends to 0, and to 0.5 for the remaining ones~\cite{Arikan2009}.

In an $(N, k)$ polar code, we use the $k$ most reliable bits to transmit the information bits and set the remaining $N-k$ bits, called the frozen bits, to zero. The location of the information and frozen bits, for an additive white Gaussian noise (AWGN) channel, can be determined using the method described in \cite{Tal2011a}. To improve the bit error rate (BER), systematic encoding can used, \cite{Arikan2011}, as is done in this work.

Polar codes are decoded using successive-cancellation decoding~\cite{Arikan2009}, which works by successively estimating a bit $\est{i}$, $i=0,...,N-1$, using the channel output $\mvec{y}$ and the previously estimated bits $\est{0}$ to $\est{i-1}$.

As shown in \cite{Leroux2013}, this can be carried out without the use of multiplications or divisions by expressing probabilities as log-likelihood-ratios (LLRs), denoted $\llr{}$, and applying the min-sum (MS) approximation. The decision rule for $\est{i}$ becomes
\begin{equation}
\label{eq:sc:ms:decision}
  \est{i} = \begin{cases}
    0,& \text{if } \llr{u_i} \geq 0;\\
    1,& \text{otherwise;}
  \end{cases}
\end{equation}
where, for $\llr{u_0}$ and $\llr{u_1}$,
\begin{equation}
\label{eq:sc:ms:f}
\llr{u_0} = f(\llr{v_0}, \llr{v_1}) = \sgn{\llr{v_0}}\sgn{\llr{v_1}} \min(|\llr{v_0}|, |\llr{v_1}|);
\end{equation}
and 
\begin{equation}
\label{eq:sc:ms:g}
\llr{u_1} = g(\llr{v_0}, \llr{v_1}, \est{0}) = \begin{cases}
\llr{v_0} + \llr{v_1} & \text{when } \est{0} = 0,\\
-\llr{v_0} + \llr{v_1} & \text{when } \est{0} = 1.\\
\end{cases}
\end{equation}

While the SC decoding algorithm of polar codes has been proven to achieve the channel capacity asymptotically in code length, it inherently has a low throughput due to the serial update of the decisions $\est{i}$.

\section{Tree Structure and Fast-SSC Decoding}
\label{sect:RSM-SCC}

\begin{figure}[t]
  \centering
  \subfloat[SC]{\label{fig:sc-tree} \input{sc-tree.tikz}}
  \quad
  \subfloat[]{\label{fig:rsm-ssc-tree} \input{rsm-ssc-tree.tikz}}
  \quad
  \subfloat[]{\label{fig:rsm-ssc-repspc} \input{rsm-ssc-repspc.tikz}}
  \caption{Decoder trees for (a) the SC decoder, and the Fast-SSC decoder (b) without and (c) with Repetition-SPC nodes.}
\end{figure}
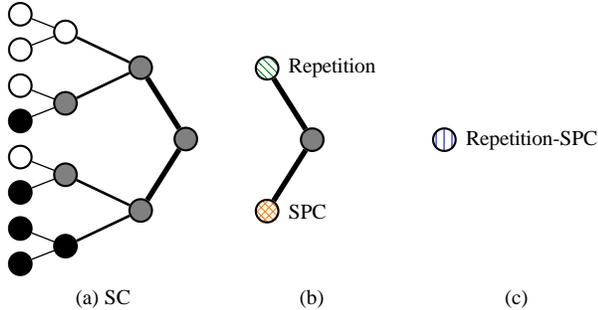

As mentioned in Section~\ref{sect:polar}, a polar code is built recursively: a polar code of length $N$ is the concatenation of two constituent polar codes of lengths $N/2$. Hence, this construction can be represented as a binary tree where each node corresponds to a constituent code \cite{Alamdar-Yazdi2011, Sarkis2013, Sarkis2013b}. Fig.~\ref{fig:sc-tree} shows the tree representation of a polar code where every node is visited, be it an information node (black) or a frozen node (white).

\subsection{Fast-SSC Decoding}

The Fast-SSC algorithm consists of several operations and is thouroughly described in \cite{Sarkis2013b}, we here focus on the key aspect of Fast-SSC decoding. It works by recognizing more types of constituent codes that, when decoded directly, significantly reduce the size and depth of the decoding tree. In this section, we briefly review these codes.

\textit{Repetition codes: }%
occur when only the last bit of a constituent code is an information bit. Shown as the node with a green striped pattern in Fig.~\ref{fig:rsm-ssc-tree}, repetition codes can be efficiently decoded by summing the input LLRs. Using threshold detection, the sign of the sum is used to determine the result, which is then replicated to form the vector of estimated bits.

\textit{Single-parity-check codes: }%
when bits of a constituent code are all information bits except the first one, it is a single-parity-check code (SPC). An SPC node is shown as a cross-hatched orange node in Fig.~\ref{fig:rsm-ssc-tree}. Such codes are decoded by first calculating the hard decision of each LLR and by calculating the parity of these decisions. The estimate of the bit with the smallest LLR magnitude is flipped when the parity constraint is unsatisfied.

\textit{Repetition-SPC codes: }%
correspond to nodes whose left child is a repetition code and the right an SPC one, shown as a node with blue vertical lines in Fig.~\ref{fig:rsm-ssc-repspc}. The speculative nature of decoding such a code is beneficial in a hardware implementation, but not in software. Therefore, in this work, the calculations for the SPC code are delayed until the decision about the repetition code is taken.

If the Fast-SSC decoding algorithm was to only recognize the repetition and SPC codes, the code of Fig.~\ref{fig:sc-tree} would be decoded in three steps as shown in Fig.~\ref{fig:rsm-ssc-tree}. Including the Repetition-SPC codes, the code is decoded in only one step as shown in Fig.~\ref{fig:rsm-ssc-repspc}.

\section{The Fast-SSC Algorithm on an x86 CPU}
\label{sect:map}
In software-defined-radio (SDR) applications, general purpose x86 CPUs are often used to carry out most of the signal processing (e.g. the Intel Core i7-2600 processor in the 2013 DARPA Spectrum Challenge). The same CPU is used to evaluate the performance of our software polar decoders based on the Fast-SSC algorithm in this work. This general-purpose x86 CPU provides support for the 128-bit Streaming SIMD Extensions (SSE) and the 256-bit Advanced Vector Extensions (AVX) and consists of four cores clocked at 3.4~GHz.

We present the results for three C++ implementations of the polar decoder that share the same memory management code, but differ in how the computational parts are implemented:
The first implementation---referred to as Float in this work---uses single-precision floating-point values without any explicit attempts at vectorization. This decoder sets the baseline for the throughput comparison in Section~\ref{sect:results}.
The second version, denoted SIMD-Float, uses the Vc C++ library \cite{Kretz2012} to perform vectorization. This enables the decoder to utilize the AVX instructions on the target platform and fall back to SSE instructions where AVX is not supported.
The third decoder uses 8-bit signed integer data and explicitly uses the 8-bit integer operations provided by the SSE extensions by means of the compiler-provided SSE intrinsics. Support for integer operations in AVX instructions is not available prior to AVX2. AVX2 is not available on the targeted i7-2600 CPU. This decoder is referred to as SIMD-int8.

\subsection{Quantization}
\begin{figure}[t]
  \centering
  \input{perf_quant.tex}
  \caption{Effect of quantization on error-correction performance of a $(32768,27568)$ polar code.}
  \label{fig:perf_quant}
\end{figure}
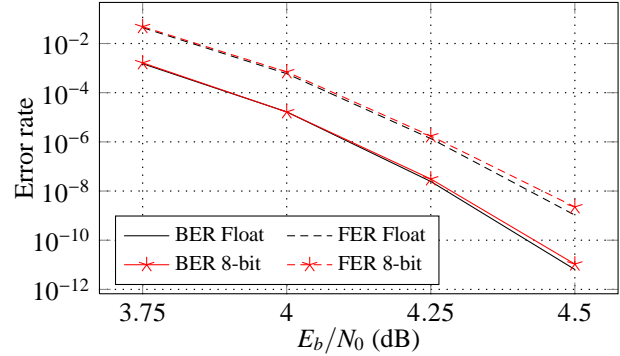

It was shown in \cite{Sarkis2013b} that, for some codes of length $32 768$, using 7 bits of quantization for LLRs results in a negligible degradation of error-correction performance over a floating-point representation. Therefore we propose that the 8-bit signed integer type used in the SIMD-int8 decoder is sufficient to achieve good error-correction performance for these codes. Figure~\ref{fig:perf_quant} confirms this assumption. At a frame-error rate (FER) of $10^{-8}$ the performance loss over floating-point is less than $0.025$~dB.

\subsection{Software Mapping to a SIMD Architecture}
The vector instructions added with SSE, up to version 4.1, support logic and arithmetic operations on vectors containing either 4 single-precision floating-point numbers or 16 8-bit integers. Additionally, our decoder uses AVX instructions, when available, to operate simultaneously on 8 packed single-precision floating-point numbers. This section lists the operations that benefited the most from explicit vectorization.

\textit{$f(\llr{a},\llr{b})$: }%
\eqref{eq:sc:ms:f} is often executed on large vectors of LLRs to prepare values for other processing nodes. The min operation and the sign calculation and assignment can all be vectorized to increase speed.

\textit{$g(\llr{a}, \llr{b}, \est{j})$: }%
This operations is also often executed on large vectors. In the Float and SIMD-Float decoders, we use $\est{j} \in \{+1, -1\}$ instead of $\{0, 1\}$. As a result, \eqref{eq:sc:ms:g} can be rewritten as
\[
g(\llr{a}, \llr{b}, \est{j}) = \llr{a}*\est{j} + \llr{b}.
\]
This removes the conditional and turns $g(\cdot)$ into a multiply-accumulate operation, which can be performed efficiently in a vectorized manner on modern CPUs. In the SIMD-int8 decoder, multiplications cannot be carried out on 8-bit integers. Thus, both possibilities of \eqref{eq:sc:ms:g} are calculated and are blended together with a mask to build the result. 

\textit{COMBINE: }%
The COMBINE operation combines two estimated bit-vectors using an XOR operation when $\est{j} \in \{0, 1\}$, or a multiplication when $\est{j} \in \{+1, -1\}$. The former is used in the SIMD-int8 decoder and the latter in the SIMD-Float decoder.

\textit{SPC Codes: }%
Locating the LLR with the minimum magnitude is accelerated using SIMD instructions.

\section{Experimental Results}
\label{sect:results}

\subsection{Methodology}
In this section, we compare the throughput, in information bits per second, of the proposed software polar decoders with that of the fastest software decoders in literature. When available, latency is also compared. The software was compiled using the C++ compiler from GCC 4.8.1 using the flags ``\texttt{-march=native -funroll-loops -Ofast}''. Additionally, auto-vectorization and link-time optimization were also enabled for all versions. The decoder is inserted in a digital communication chain to measure its performance. We use binary phase shift keying (BPSK) over an AWGN channel with random codewords.

The throughput is calculated using the time required to decode a frame averaged over 10 runs of $\num[group-separator={,}]{50000}$ and $\num[group-separator={,}]{10000}$ frames each for the $N=2048$ and the $N>2048$ codes, respectively. The time required to decode a frame, or latency, also includes the time required to copy a frame to decoder memory and copy back the estimated codeword, and is measured using the high precision clock provided by the Boost Chrono library. Codeword generation, transmission over the channel, demodulation, and, for SIMD-int8, quantization are excluded from calculations.

\subsection{Comparison of the Software Implementations}
\label{sect:cmp_pc}
As shown in Table~\ref{tab:impl:ml-ssc:tp}, the vectorized implementations are 3 to 5 times faster than the Float decoder for the $(32768,27568)$ and $(32768,29492)$ codes. Table~\ref{tab:impl:ldpc} shows that for the shorter $N=2048$ codes, the speedup factors are similar at approximately 3 for SIMD-float and greater than 4 for SIMD-int8.
\begin{table}
  \centering
  \caption{Throughput comparison of decoding polar codes of length $N=\num[group-separator={,}]{32768}$ using Fast-SSC.}
  \resizebox{0.5\textwidth-0.5\columnsep}{!}{%
  \begin{tabular}{c c c c c c}
    \toprule
    \multirow{2}{*}{Code rate} & \multirow{2}{*}{Implementation} & \multicolumn{2}{c}{T/P (Mbps)} & \multirow{2}{*}{\shortstack{Latency \\ ($\mu$s)}} \\
    \cmidrule{3-4}
                   &                               & Coded & Info & & \\
    \midrule
    0.84 & Float      & 47.47 & 39.93 & 690\\
         & SIMD-Float & 147.01 & 123.68 & 223\\
         & SIMD-int8  & 242.01 & 203.60 & 135\vspace{2pt}\\
    0.9  & Float      & 54.04 & 48.64 & 606\\
         & SIMD-Float & 173.77 & 156.40 & 189\\
         & SIMD-int8  & 252.06 & 226.86 & 130\\
    \bottomrule
  \end{tabular}%
  }
  \label{tab:impl:ml-ssc:tp}
\end{table}

\subsection{Comparison with Software LDPC Decoders}
Software LDPC decoders are presented in \cite{Falcao2009,GPUFalcao2009,Falcao2011,GPULi2013,GPUWang2013}. The decoders of \cite{GPUFalcao2009,Falcao2011} target the Cell/BE multicore processor, an NVIDIA 8800 GTX GPU and an Intel x86 CPU, respectively. In \cite{Falcao2009}, Falc\~{a}o et al. cover more WiMAX LDPC codes with their implementation for the Cell/BE processor. Lastly, \cite{GPULi2013} and \cite{GPUWang2013} are aimed at delivering very high throughput using modern GPUs. In all cases, these software LDPC decoders parallelize the decoding of multiple received frames whereas we parallelize the decoding of a single frame. If the proposed polar decoders use all four cores of the CPU to simultaneously decode 4 frames, the throughput approximately quadruples. However in typical SDR applications the remaining cores are used for other tasks such as demodulation.

We focus on the moderate length LDPC codes, with rates $1/2$ and $5/6$, from the WiMAX standard, which are used in \cite{Falcao2009,GPUFalcao2009,Falcao2011,GPULi2013,GPUWang2013}.
We compare software LDPC and polar decoders at a similar error-correction performance for a given rate. Thus, the polar codes chosen for the comparison were selected accordingly.
The $(2048, 1024)$ and $(2048,1707)$ polar codes match the $(2304, 1152)$ and $(1248,1040)$ LDPC codes, respectively, decoded using the min-sum algorithm with 10 iterations.

\begin{figure}[t]
  \centering
  \input{perf_vs_ldpc.tex}
  \caption{Error-correction performance of polar codes compared with that of LDPC codes with the same rate.}
  \label{fig:perf_vs_ldpc}
\end{figure}
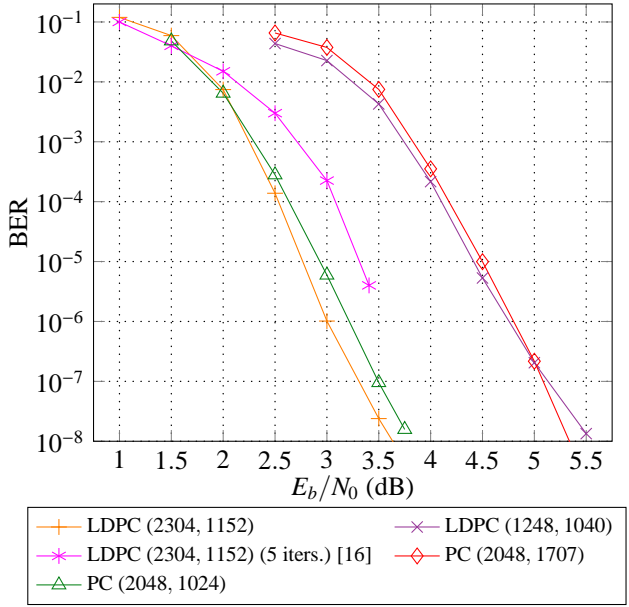

Fig.~\ref{fig:perf_vs_ldpc} shows the error-correction performance of these codes. It can be seen that the $(2048, 1707)$ polar code was about 0.05~dB away from the longest rate-$5/6$ code of \cite{Falcao2009}. The error-correction performance of the $(2048, 1024)$ polar code was 0.2~dB worse than that of the $(2304, 1152)$ LDPC code when decoded with 10 iterations, but better than the LDPC decoder using only 5 iterations.

Table~\ref{tab:impl:ldpc} shows the information throughput and latency corresponding to the proposed polar decoders in comparison with that of 10 decoding iterations of the software LDPC decoders of \cite{Falcao2009,GPUFalcao2009,Falcao2011,GPUWang2013}. The LDPC decoder of \cite{GPULi2013} is also shown but only uses 5 iterations. For the highest rate $5/6$ code, both of our proposed decoders using SIMD instructions have a better throughput than the software LDPC decoder even though we only use one CPU core. For the codes with rate $1/2$, we would need to use all four cores of our CPU in order to surpass the throughput of the fastest LDPC decoder on GPU.

As shown in Table~\ref{tab:impl:ldpc}, the latency of the SIMD polar decoders, at 10--14~$\mu$s, is an order of magnitude smaller than that of the GPU and Cell/BE implementations of \cite{GPUFalcao2009,Falcao2011,GPUWang2013}. This is due in part to the LDPC decoders buffering multiple frames, e.g. 16 for \cite{GPUFalcao2009} and 50 for \cite{GPUWang2013}.

It should be noted that the implementations of \cite{Falcao2009,GPUFalcao2009,Falcao2011,GPUWang2013,GPULi2013} trade error-correction performance for throughput and that these LDPC codes can perform better when more iterations are used. For example, the use of 5 decoding iterations in \cite{GPULi2013} instead of 10 leads to an error-correction performance degradation of 0.5~dB, as shown in Fig.~\ref{fig:perf_vs_ldpc}.

\begin{table}[t]
  \centering
  \caption{Comparison with software LDPC decoders for codes of rates 1/2 and 5/6.}
  \resizebox{0.5\textwidth-0.5\columnsep}{!}{%
  \begin{tabular}{c c c C{1.2cm} c}
    \toprule
    \multirow{2}{*}{Decoder} & \multirow{2}{*}{$(N,k)$} & \multirow{2}{*}{Target} & Info T/P & \multirow{2}{*}{\shortstack{Latency \\ ($\mu$s)}} \\
                          &                    &                               & (Mbps) &\\
    \midrule
    LDPC\cite{Falcao2011} & $(1024,512)$       & x86 CPU    & 1.04 & N/A\vspace{3pt}\\
    LDPC\cite{GPUFalcao2009} & $(1024,512)$    & GPU        & 20.9 & 393\vspace{3pt}\\
    LDPC\cite{Falcao2011} & $(1024,512)$       & Cell/BE    & 34.8 & 354\vspace{3pt}\\
    LDPC\cite{GPUWang2013}& $(2304,1152)$      & GPU        & 152.1 & 1266\vspace{3pt}\\
    LDPC\cite{GPULi2013} (5 iters.) & $(2304,1152)$ & GPU   & 355.0 & N/A\vspace{3pt}\\
    PC Float              & $(2048,1024)$      & x86 CPU    & 23.2 & 44\\
    PC SIMD-Float         &                    &            & 71.5 & 14\\
    PC SIMD-int8          &                    &            & 101.7 & 10\\
    \midrule

    LDPC\cite{Falcao2009} & $(1248,1040)$      & Cell/BE    & 65.3 & N/A\vspace{3pt}\\
    PC Float              & $(2048,1707)$      & x86 CPU    & 45.4 & 38\\
    PC SIMD-Float         &                    &            & 154.1 & 12\\
    PC SIMD-int8          &                    &            & 198.7 & 9\\
    \bottomrule
  \end{tabular}%
  }
  \label{tab:impl:ldpc}
\end{table}

\subsection{A Note About Hardware SC Polar Decoders}
\label{sect:cmp_hwsc}
In terms of throughput, the proposed software decoders are competitive with all hardware decoders in literature, with the exception of the hardware implementation of the same (Fast-SSC) algorithm \cite{Sarkis2013b}. For example, despite using a smaller number of quantization bits, the semi-parallel polar decoder of \cite{Leroux2013} offers an inferior throughput for the $(2048,1707)$ and $(32768,29462)$ codes, achieving only 69.2~Mbps and 27.6~Mbps respectively, whereas the SIMD-int8 decoder reaches 198.7~Mbps and 226.9~Mbps. The decoder is also faster than the hardware polar decoders of \cite{Raymond2013,TWOPhasePamuk}. In the case of the latter, the two phase SC decoder~\cite{TWOPhasePamuk}, the fastest non-Fast-SSC decoder in literature, has an information throughput of 102.6~Mbps for a $(16384,14746)$ code. Our 8-bit SIMD decoder achieves 242.3~Mbps for the same code.

\section{Conclusion}
\label{sect:conclusion}
In this work, we presented fast software implementations of polar decoders. By taking advantage of the SIMD extensions of a common x86 CPU, our implementation was able to achieve an information throughput greater than 200~Mbps by only using one CPU core clocked at 3.4~GHz. Moreover, for polar codes with similar error-correction performance compared to that of LDPC codes, we are able to obtain a greater throughput than LDPC decoders targeting x86 CPUs and the Cell/BE processor; and is competitive with state-of-the-art software GPU-based decoders. In addition, this software decoder is faster than any hardware polar decoder with exception of the one implementing the same algorithm.

Finally, our initial experiments with an Intel Haswell processor core, featuring the AVX2 supplementary instructions, gave us an information throughput greater than 300~Mbps for a core clocked at 3.4~GHz by using the new instructions operating simultaneously over 32 8-bit integers packed in a 256-bit register. Hence, our results indicate that polar codes are promising candidates for software-defined-radio applications.

\section*{ACKNOWLEDGEMENT}
The authors wish to thank Alexandre J. Raymond and Fran\c{c}ois Leduc-Primeau of McGill University for helpful discussions. Claude Thibeault is a member of ReSMiQ.

\bibliographystyle{IEEEtran}
\bibliography{IEEEabrv,simd_mlssc.bib}

\end{document}

%% file: sc-tree.tikz
\begin{tikzpicture}[baseline = (0_7.center),
        level/.style={level distance = 6mm},
        level 1/.style={sibling distance=19mm, edge from parent/.style={draw,black,line width=2pt}},
        level 2/.style={level distance=10mm, sibling distance=9.5mm, edge from parent/.style={draw,black,line width=1pt}},
        level 3/.style={sibling distance=4.7mm, edge from parent/.style={draw,black,line width=0.5pt}},
        ]

\tikzset{
frozen/.style={thick,draw=black,fill=white,minimum size=3mm,circle, inner sep=0},
fullspace/.style={thick,draw=black,fill=black,minimum size=3mm,circle, inner sep = 0},
mixed/.style={thick,draw=black,fill=gray,minimum size=3mm,circle, inner sep = 0},
ml_mixed/.style={thick,draw=black,fill=blue,minimum size=3mm,circle, inner sep = 0}
}

\node[mixed] (p){} [grow=left]
	child {node[mixed] (2_0){}
		child {node[frozen] (1_0){}
			child {node[frozen] (a0_0){}
			}
			child {node[frozen] (a0_1){}
			}
		}
		child {node[mixed] (1_2){}
			child {node[frozen] (0_2){}
			}
			child {node[fullspace] (0_3){}
			}
		}
	}
	child {node[mixed] (v){}
		child {node[mixed] (cl){}
			child {node[frozen] (0_4){}
			}
			child {node[fullspace] (0_5){}
			}
		}
		child {node[fullspace] (cr){}
			child {node[fullspace] (0_6){}
			}
			child {node[fullspace] (0_7){}
			}
		}
	}
;

\end{tikzpicture}

%% file: rsm-ssc-tree.tikz
\definecolor{deepgreen}{RGB}{8, 130, 25}

\begin{tikzpicture}[baseline=(base),
        level/.style={level distance = 6mm},
        level 1/.style={sibling distance=19mm, edge from parent/.style={draw,black,line width=2pt}},
        level 2/.style={sibling distance=9mm, edge from parent/.style={draw,black,line width=1pt}},
        level 3/.style={sibling distance=4mm, edge from parent/.style={draw,black,line width=0.5pt}},
        ]

\tikzset{
frozen/.style={thick,draw=black,fill=white,minimum size=3mm,circle, inner sep=0},
fullspace/.style={thick,draw=black,fill=black,minimum size=3mm,circle, inner sep = 0},
mixed/.style={thick,draw=black,fill=gray,minimum size=3mm,circle, inner sep = 0},
rep_mixed/.style={thick,draw=black,pattern=north west lines,pattern color=deepgreen,minimum size=3mm,circle, inner sep = 0},
spc_mixed/.style={thick,draw=black,pattern=crosshatch,pattern color=orange,minimum size=3mm,circle, inner sep = 0},
repspc/.style={thick,draw=black,pattern=vertical lines,pattern color=blue,minimum size=3mm,circle, inner sep = 0}
}

\node[mixed] (3_0){} [grow=left]
	child {node[rep_mixed,label={[label distance=0cm]2:{\footnotesize Repetition}}] (2_0){}
	}
	child {node[spc_mixed,label={[label distance=0cm]-2:{\footnotesize SPC}}] (2_1){}
	}
;
	
\node [circle, below= 5.27mm of 2_1.base] (base) {};
\node[circle, left= 4mm of 3_0] (pad1) {};
\end{tikzpicture}

%% file: rsm-ssc-repspc.tikz
\definecolor{deepgreen}{RGB}{8, 130, 25}

\begin{tikzpicture}[baseline=(base),
        level/.style={level distance = 6mm},
        level 1/.style={sibling distance=19mm, edge from parent/.style={draw,black,line width=2pt}},
        level 2/.style={sibling distance=9mm, edge from parent/.style={draw,black,line width=1pt}},
        level 3/.style={sibling distance=4mm, edge from parent/.style={draw,black,line width=0.5pt}},
        ]

\tikzset{
frozen/.style={thick,draw=black,fill=white,minimum size=3mm,circle, inner sep=0},
fullspace/.style={thick,draw=black,fill=black,minimum size=3mm,circle, inner sep = 0},
mixed/.style={thick,draw=black,fill=gray,minimum size=3mm,circle, inner sep = 0},
rep_mixed/.style={thick,draw=black,pattern=north west lines,pattern color=deepgreen,minimum size=3mm,circle, inner sep = 0},
spc_mixed/.style={thick,draw=black,pattern=crosshatch,pattern color=orange,minimum size=3mm,circle, inner sep = 0},
repspc/.style={thick,draw=black,pattern=vertical lines,pattern color=blue,minimum size=3mm,circle, inner sep = 0}
}

\node[repspc, label={0:{\footnotesize Repetition-SPC}}] (4_0){};

\node [circle, below= 14.8mm of 4_0.base] (base) {};

\end{tikzpicture}

%% file: perf_quant.tex
\begin{tikzpicture}

\pgfplotsset{
  grid style = {
    dash pattern = on 0.05mm off 1mm,
    line cap = round,
    black,
    line width = 0.5pt
  }
}

\begin{semilogyaxis}[%
    xlabel=$E_b/N_0$ (dB),xtick={3.75,4.00,4.25,4.5},%
    xlabel style={yshift=0.6em},%
    ylabel=Error rate, ylabel style={yshift=-0.65em},%
    ytick={1e-2,1e-4,1e-6,1e-8,1e-10,1e-12},%
    width=\textwidth/2.10, height=5.4575cm, grid=major,%
    legend pos=south west,%
    legend style={/tikz/every even column/.append style={column sep=0.22cm, font=\footnotesize}},%
    legend cell align=left, legend columns=2,%
    mark size=3.0pt]

\addplot[color=black] coordinates {
  (3.75000e+00, 1.50819e-03)
  (4.00000e+00, 1.65693e-05)
  (4.25000e+00, 2.44926e-08)
  (4.50000e+00, 6.446041e-12)
};
\addlegendentry{BER Float}
\addplot[color=black,densely dashed] coordinates {
  (3.75000e+00, 4.42159e-02)
  (4.00000e+00, 5.93023e-04)
  (4.25000e+00, 1.35855e-06)
  (4.50000e+00, 1.049953e-09)
};
\addlegendentry{FER Float}

\addplot[color=red,mark=star] coordinates {
  (3.75000e+00, 1.64072e-03)
  (4.00000e+00, 1.65096e-05)
  (4.25000e+00, 3.09430e-08)
  (4.50000e+00, 1.06652e-11)
};
\addlegendentry{BER 8-bit}

\addplot[color=red,densely dashed,mark=star] coordinates {
  (3.75000e+00, 4.83178e-02)
  (4.00000e+00, 7.00902e-04)
  (4.25000e+00, 1.73600e-06)
  (4.50000e+00, 2.25751e-09)
};
\addlegendentry{FER 8-bit}
\end{semilogyaxis}
\end{tikzpicture}

%% file: perf_vs_ldpc.tex
\definecolor{deepgreen}{RGB}{8, 130, 25}
\definecolor{mauve}{RGB}{152, 57, 153}
\definecolor{hotpink}{RGB}{255, 0, 255}

\begin{tikzpicture}

\pgfplotsset{
  grid style = {
    dash pattern = on 0.05mm off 1mm,
    line cap = round,
    black,
    line width = 0.5pt
  }
}

\begin{semilogyaxis}[xlabel=$E_b/N_0$ (dB), xlabel style={yshift=0.55em},%
                     xtick={0,1,1.5,2,2.5,3,3.5,4,4.5,5,5.5},
                     ylabel=BER, ylabel style={yshift=-0.65em},%
                     ytick={1e-1,1e-2,1e-3,1e-4,1e-5,1e-6,1e-7,1e-8},%
                     ymax=2e-1,ymin=1e-8,%
                     xmin=0.75,xmax=5.75,%
                     width=\textwidth/2.10, height=7.4cm,%
                     grid=major,%
                     legend style={/tikz/every even column/.append
                     style={column sep=0.22cm,
                     font=\footnotesize}, at={(0.45,-0.15)},
                     anchor=north},%
                     legend cell align=left,%
                     legend columns=2, mark size=3.0pt]

\addplot[color=orange,mark=+] coordinates {
  (1.00000e+00, 1.19216e-01)
  (1.50000e+00, 5.92025e-02)
  (2.00000e+00, 7.43609e-03)
  (2.50000e+00, 1.38124e-04)
  (3.00000e+00, 1.01531e-06)
  (3.50000e+00, 2.39114e-08)
  (3.75000e+00, 4.73264e-09)
};
\addlegendentry{LDPC $(2304, 1152)$}

\addplot[color=mauve, mark=x] coordinates {
  (2.50000e+00, 4.35887e-02)
  (3.00000e+00, 2.26736e-02)
  (3.50000e+00, 4.27611e-03)
  (4.00000e+00, 2.15859e-04)
  (4.50000e+00, 5.31808e-06)
  (5.00000e+00, 2.0456e-07)
  (5.50000e+00, 1.34011e-08)
};
\addlegendentry{LDPC $(1248, 1040)$}

\addplot[color=hotpink,mark=asterisk] coordinates {
  (1.00000e+00, 1.0e-01)
  (1.50000e+00, 4.0e-02)
  (2.00000e+00, 1.5e-02)
  (2.50000e+00, 3.0e-03)
  (3.00000e+00, 2.25e-04)
  (3.40625e+00, 4.0e-06)
};
\addlegendentry{LDPC $(2304, 1152)$ (5 iters.)~\cite{GPULi2013}}

\addplot[color=red,mark=diamond] coordinates {
  (2.50000e+00, 6.55440e-02)
  (3.00000e+00, 3.75637e-02)
  (3.50000e+00, 7.47982e-03)
  (4.00000e+00, 3.52802e-04)
  (4.50000e+00, 9.95001e-06)
  (5.00000e+00, 2.16750e-07)
  (5.50000e+00, 2.29634e-09)
};
\addlegendentry{PC $(2048, 1707)$}

\addplot[color=deepgreen,mark=triangle] coordinates {
  (1.50000e+00, 4.82789e-02)
  (2.00000e+00, 6.47338e-03)
  (2.50000e+00, 2.80434e-04)
  (3.00000e+00, 5.94796e-06)
  (3.50000e+00, 9.53237e-08)
  (3.75000e+00, 1.57626e-08)
};
\addlegendentry{PC $(2048, 1024)$}

\end{semilogyaxis}
\end{tikzpicture}